\documentclass[twocolumn,letterpaper,aps,pra,superscriptaddress,showpacs,amsmath]{revtex4-1}

\newcommand{\bea}{\begin{eqnarray}}
\newcommand{\eea}{\end{eqnarray}}
\newcommand{\beq}{\begin{equation}}
\newcommand{\eeq}{\end{equation}}

\usepackage[urlcolor=blue,colorlinks=true,citecolor=blue,linkcolor=blue,pdfstartview={FitH},bookmarks=false]{hyperref}
\usepackage{graphicx}
\usepackage{longtable}
\usepackage{epsfig}
\usepackage{dcolumn}
\usepackage{bm}
\usepackage{amssymb}
\usepackage{multirow}
\usepackage{times,color}
\usepackage{hyperref}
\usepackage{braket}
\usepackage{comment}
\begin{document}

\title{Character of Doped Holes in Nd$_{1-x}$Sr$_x$NiO$_2$}


\author{Tharathep Plienbumrung}
\affiliation{\mbox{Institute for Functional Matter and Quantum Technologies,
University of Stuttgart, Pfaffenwaldring 57, D-70550 Stuttgart, Germany}}
\affiliation{\mbox{Center for Integrated Quantum Science and Technology,
University of Stuttgart, Pfaffenwaldring 57, D-70550 Stuttgart, Germany}}

\author{Michael Schmid}
\affiliation{\mbox{Waseda Research Institute for Science and Engineering,
Waseda University, Okubo, Shinjuku, Tokyo, 169-8555, Japan}}

\author{Maria Daghofer}
\affiliation{\mbox{Institute for Functional Matter and Quantum Technologies,
University of Stuttgart, Pfaffenwaldring 57, D-70550 Stuttgart, Germany}}
\affiliation{\mbox{Center for Integrated Quantum Science and Technology,
University of Stuttgart, Pfaffenwaldring 57, D-70550 Stuttgart, Germany}}

\author{Andrzej M. Ole\'s$\,$}
\email{a.m.oles@fkf.mpi.de}
\affiliation{\mbox{Max Planck Institute for Solid State Research,
             Heisenbergstrasse 1, D-70569 Stuttgart, Germany} }
\affiliation{\mbox{Institute of Theoretical Physics, Jagiellonian University,
             Profesora Stanis\l{}awa \L{}ojasiewicza 11, PL-30348 Krak\'ow, Poland}}

\begin{abstract}
We investigate charge distribution in the recently discovered 
high-$T_c$ superconductors, layered nickelates. With increasing value 
of charge-transfer energy we observe the expected crossover from the 
cuprate to the local triplet regime upon hole doping. We find that 
the $d-p$ Coulomb interaction $U_{dp}$ plays a role and makes Zhang-Rice
singlets less favorable, while the amplitude of local triplets is 
enhanced. By investigating the effective two-band model with orbitals of 
$x^2-y^2$ and $s$ symmetries we show that antiferromagnetic interactions 
dominate for electron doping. The screened interactions for the $s$ band 
suggest the importance of rare-earth atoms in superconducting nickelates.
\end{abstract}

\date{29 July, 2021}

\maketitle


\section{Introduction: Superconducting infinite-layered nickelates
N\lowercase{d}N\lowercase{i}O$_2$}

The discovery of Bednorz and Muller \cite{Bed86} started intense search 
for novel superconductors with high values of the critical temperature 
$T_c$. But in spite of a tremendous effort in the theory, the mechanism 
responsible for the pairing in cuprates is still unknown \cite{Kei15}.
Yet, this is one of the fundamental open problems in modern physics.

Perhaps less spectacular was the recent discovery of superconductivity 
in infinite-layered NdNiO$_{2}$ doped by Sr \cite{Li19} as the values of 
$T_c$ are "only" close to 15 K \cite{Li20}. Nevertheless, it gave a new 
impulse to the theory of high-$T_c$ superconductivity at large. To some 
extent, the nickelate superconductor family is rather similar to cuprate 
superconductors \cite{Bot20} as once again two-dimensional (2D) planes 
of transition metal and oxygen ions play a central role here \cite{Bia19}. 
With Ni$^{+}$ ions one has again $d^9$ electronic configuration and similar 
lattice structure but the apical oxygens are absent. By following the 
same analysis for a NiO$_2$ layer as one performed for a CuO$_2$ layer, 
we expect Ni$^{1+}$ to exhibit antiferromagnetic (AFM) order. One finds
that the charge-transfer gap in nickelate is larger than that of
cuprate \cite{Jia20,Tha21}. The doped holes reside on oxygen sites in
cuprates forming the Zhang-Rice singlet \cite{Zha88}. On the contrary,
the doped holes will likely reside on Ni sites in doped 
Nd$_{1-x}$Sr$_x$NiO$_2$.

\section{Charge-transfer model: N\lowercase{i}O$_2$ model revisited}

We introduced the multiband $d-p$ Hamiltonian for a NiO$_2$ plane 
\cite{Tha21} starting from a 2D Ni$_4$O$_8$ $2\times 2$ cluster with 
periodic boundary conditions (PBCs). The basis set includes four 
orbitals per NiO$_2$ unit cell: two $e_g$ orbitals 
\mbox{$\{3z^2-r^2,x^2-y^2\}\equiv \{z,\bar{z}\}$} at each Ni$^+$ ion and 
one bonding $2p_\sigma$ orbital (either $2p_x$ or $2py$) at each oxygen 
ion in the 2D plane,
\begin{equation}
{\cal H}=H_{dp}+H_{pp}+H_{\rm diag}+H_{\rm int}^d+H_{\rm int}^p.
\label{model}
\end{equation}
Here the first two terms in the Hamiltonian (\ref{model}) stand for the
kinetic energy: $H_{dp}$ includes the $d-p$ hybridization
$\propto t_{pd}$ and $H_{pp}$ includes the interoxygen $p-p$ hopping
$\propto t_{pp}$,
\begin{eqnarray}
H_{dp}&=&\sum_{\{m\alpha;j\nu\},\sigma}\left(t_{pd}
 \hat{d}^{\dagger}_{m\alpha\sigma}\hat{p}_{j\nu\sigma}^{} + {\rm H.c.}\right),\\
H_{pp}&=&\sum_{\{i\mu;j\nu\},\sigma}\left(t_{pp}
 \hat{p}^{\dagger}_{i\mu\sigma}\hat{p}_{j\nu\sigma}^{} + {\rm H.c.}\right),
\end{eqnarray}
where $\hat{d}_{m\alpha\sigma}^{\dagger}$ ($\hat{p}_{j\nu\sigma}^{\dagger}$) 
is the creation operator of an electron at nickel site $m$ (oxygen 
site $i$) in an orbital $\alpha\in\{z,\bar{z}\}$ ($\mu\in\{x,y\}$). Here 
$z$ and $\bar{z}$ stand for $3z^2-r^2$ and $x^2-y^2$ orbitals, while 
$\{x,y\}$ stand for $p_x$ and $p_y$ orbital. The elements 
$\{t_{pd},t_{pp}\}$ are accompanied by the phase factors which follow 
from orbital phases \cite{Dag09}.

The one-particle (level) energies are included in $H_{\rm diag}$, where
we introduce the charge-transfer energy between $d$ and $p$ orbitals,
\begin{equation}
\Delta=\varepsilon_d-\varepsilon_p.
\label{Delta}
\end{equation}
The model is completed by the Coulomb interactions in the $d$ and $p$ 
orbitals. For the $d$ electrons,
\begin{eqnarray}
	\label{int}
H_{\rm int}^{d}\!&=& \sum_{m\alpha}U_{\alpha}
n_{m\alpha\uparrow}n_{m\alpha\downarrow}
+ \left(U'-\frac{1}{2}J_H\right)\sum_{i}n_{m1}n_{m2}   \nonumber \\ 
\!&-&\!2J_H\!\sum_{i}\!\vec{S}_{m1}\!\cdot\!\vec{S}_{m2}
+\!J_{H}\!\sum_{i}\!d^{\dagger}_{m1\uparrow}d^{\dagger}_{m1\downarrow}d_{m2\uparrow}^{}d_{m2\downarrow}^{},
\end{eqnarray}
where $\alpha=z,\bar{z}$ stands for $z^2$ and $x^{2}-y^2$ orbitals. 
$d^{\dagger}_{m\alpha\sigma}$ ($d_{m\alpha\sigma}$) is electron creation 
(annihilation) operator at site $m$ in orbital $\alpha$ with spin 
$\sigma$. $n_{m\alpha\sigma}$ is a number operator. $\vec{S}_{m\alpha}$ 
stands for spin operator in $\alpha$ orbital at site~$m$. 
The two Kanamori parameters to describe the interactions between $3d$ 
electrons are $\{U,J_H\}$. $U_\alpha$ is Coulomb repulsion element for 
$\alpha$ orbital. $J_H$ and $U'=U-2J_H$ are Hund's exchange and 
interorbital Coulomb interaction \cite{Ole83}. 
Similar interactions with $\{U^p,J_H^p\}$ are written for $p$ electrons.

\begin{table}[b!]
\caption{Parameters of the NiO$_2$ charge-transfer model (all in eV)
		 used in \cite{Tha21}.}
 \centering
\begin{ruledtabular}
	\begin{tabular}{ccccccc}
$t_{pd}$ & $t_{pp}$ & $\Delta$ & $U_z=U_{\bar z}$ & $J_H$ & $U_p$ & $J_H^p$\\
			\hline
 1.30 & 0.55 & $7.0$ & 8.4 & 1.2 & 4.4 & 0.8 \\
		\end{tabular}
\end{ruledtabular}
		\label{tab1}
\end{table}

\begin{figure*}[t!]
	\centering
	\includegraphics[width=14cm]{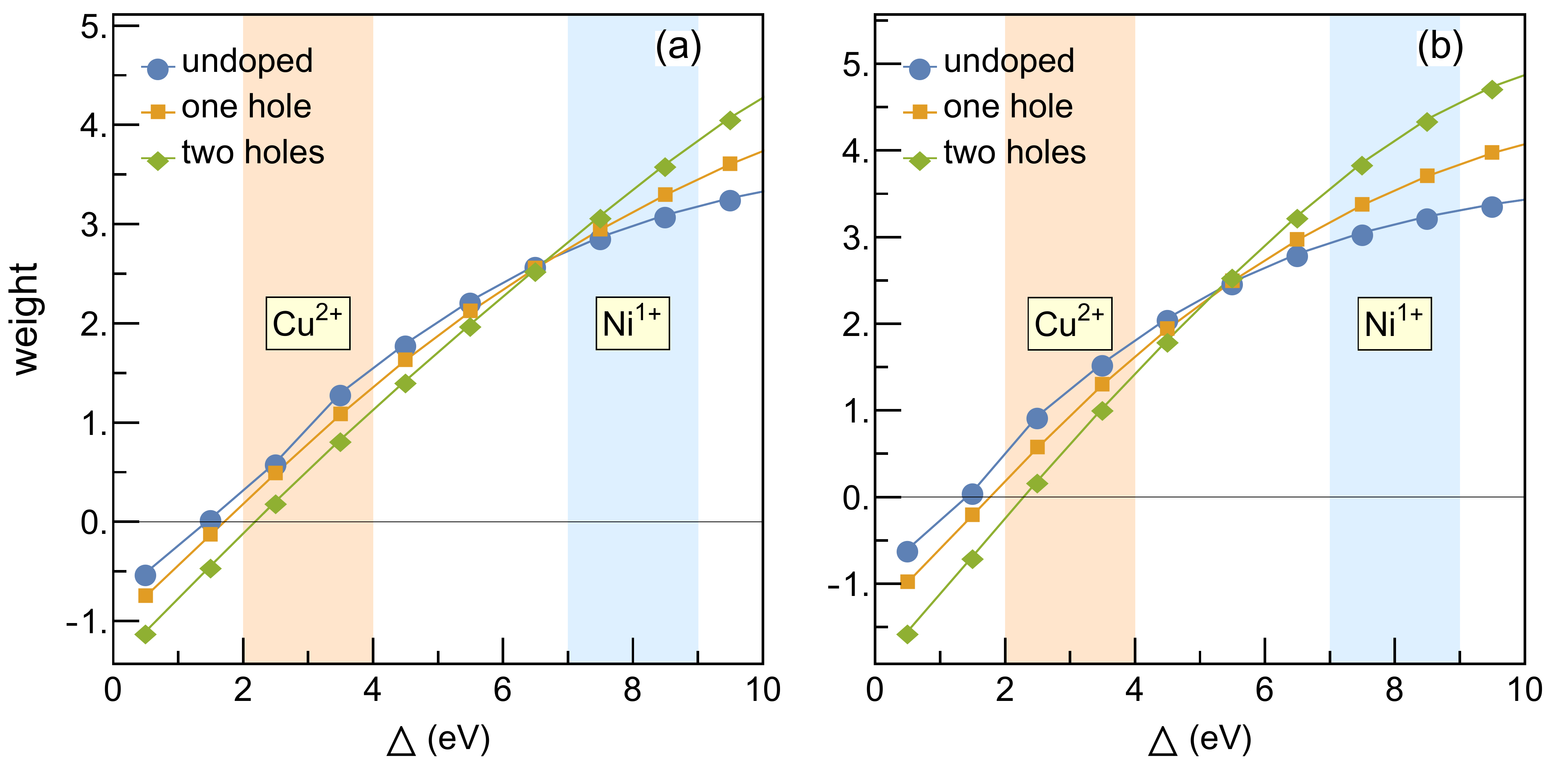}
\caption{Doping tendency of NiO$_2$ model as a function of crystal-field 
splitting $\Delta$: 
(a) $U_{dp} = 0$; 
(b)~$U_{dp} = 1$ eV. 
(Blue) undoped, $N_{\uparrow}=2$ and $N_\downarrow=2$; 
(Orange) one hole doped, $N_{\uparrow}=3$, $N_\downarrow=2$; 
(Green) two holes doped, $N_{\uparrow}=3$, $N_\downarrow=3$ 
in Ni$_4$O$_8$ (Cu$_4$O$_8$) cluster.
The nickelate (cuprate) regime is highlighted in blue (orange). 
The model parameters are given in Table \ref{tab1}. }
	\label{fig1}
\end{figure*}

Figure \ref{fig1} shows different weight distributions for hole 
occupations at Ni and O sites for the parameters of Table \ref{tab1}. 
The Ni-O charge-transfer model has been studied via impurity 
\cite{Jia20} as well as lattice approach \cite{Tha21}. It has been 
established that the holes in undoped compounds remain within 
$d_{x^2-y^2}$ orbitals. Thereby we have assumed that Nd does not 
contribute to the electronic structure and the system without Sr is a 
Mott insulator. Doping by Sr gives a doped hole which tends to reside 
at nickel sites rather than at oxygen sites. 
The nickelate (cuprate) regime \cite{Jia20} is highlighted in blue 
(orange) in Fig. \ref{fig1}. In the nickelate regime the holes reside 
predominantly at Ni sites. This is the essential difference with 
cuprates where a doped hole (for hole doping) resides predominantly at 
oxygen and forms a Zhang-Rice singlet \cite{Zha88}. 

Including intersite Coulomb repulsion $U_{dp}$ favors the hole 
occupancy at Ni sites \cite{Tha21} and shifts the doping crossover to 
lower values of the charge transfer energy $\Delta$. All the on-site 
energies of the Ni($3d$) orbitals have been included in the Ni-O 
hybridization terms $t_{pd}$~\cite{Jia20}. Similar results were 
obtained for finite $e_g$ orbital splitting, where $\Delta_z=1$ eV 
should be considered the upper limit. 

The NiO$_2$ compound is a Mott-Hubbard 
insulator. It is then possible to replace the charge-transfer model by 
the $d$-only Hubbard model. The effective Ni-Ni hoppings can be derived 
from second-order perturbation theory \cite{Hu19}. The next question is 
on which $d$-orbitals the doped holes are?

The asymmetric distribution of holes suggests that one could replace 
the Ni-O model \eqref{model} with Ni $d$-only 
model as oxygen $p$-orbitals become unimportant. The DFT calculation 
shows the band structure of NdNiO$_2$ that two bands are crossing Fermi 
level. In the orbital-resolved band structure, the lower band has 
$d_{x^2-y^2}$ character and the upper band contains both Nd and Ni 
contributions. The large charge transfer energy as well as the presence 
of electron pocket at $\Gamma$ are the two striking features of the 
nickelate compound. The empty 5d states of Nd are responsible for 
providing electron pocket. The empty $5d$ states are below the Fermi 
level, in other words, these states provide the hole states into Ni 
band by the so-called 'self-doped' effect 
\cite{ZYZ20}. Furthermore, the $5d$ states was shown to be hybridized 
with Ni apical orbitals i.e., $3d_{z^2}$ and 4$s$ orbitals \cite{Adh20} 
lead us to construct the effective two-band model consisting of Ni 
in-plane orbital, $d_{x^2- y^2}$, and Ni off-plane orbital, the 
modified $s$ orbital. In this work, we present the character of 
doped holes in the realistic two-band model of NdNiO$_2$ compound.
 
\section{Electronic structure calculations}

\begin{figure*}[t!]
	\centering
	\includegraphics[width=15cm]{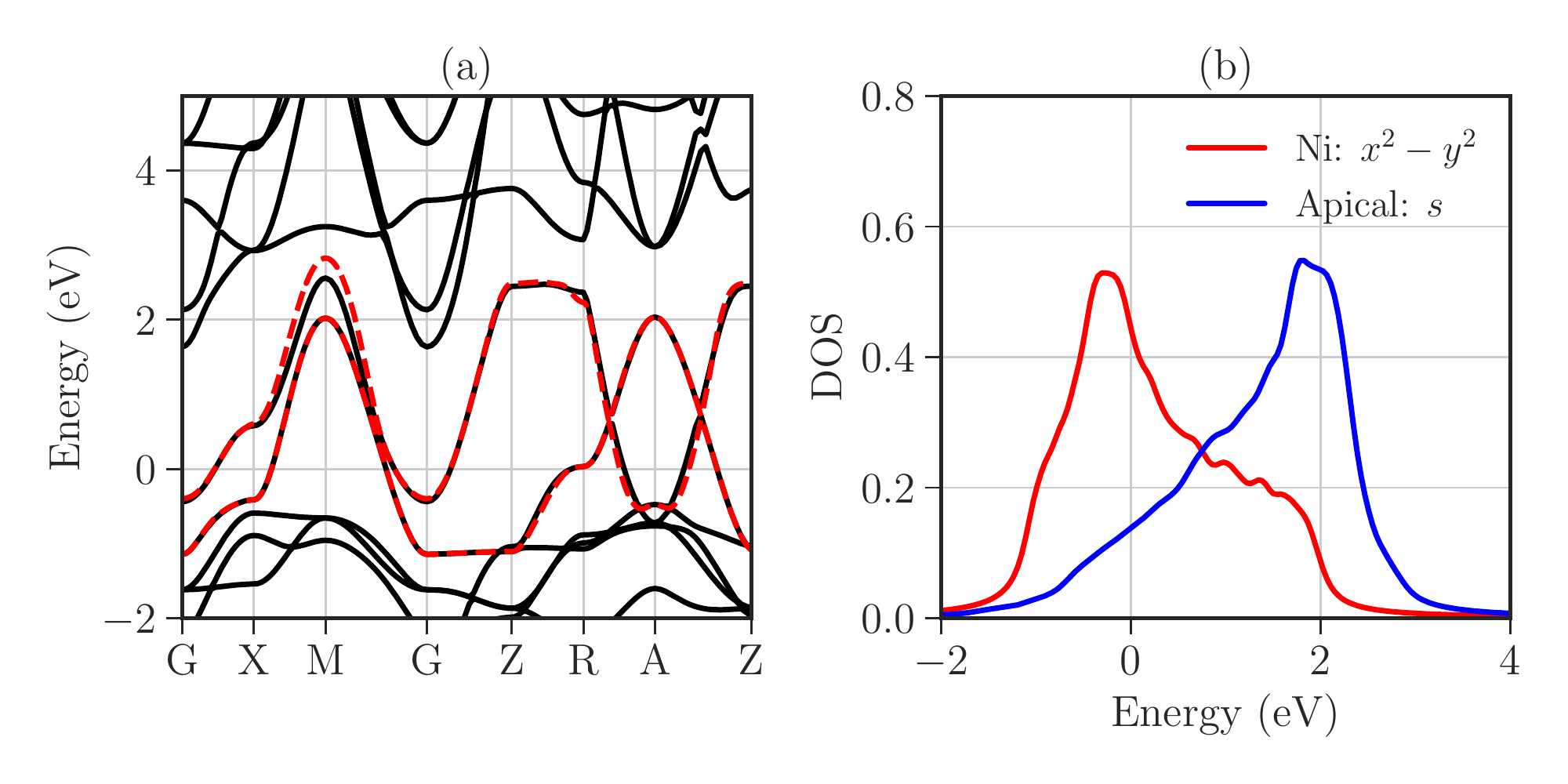}
\caption{DFT band structure and DOS of the Wannier Hamiltonian: 
(a) DFT band structure (black solid lines) compared with Wannier band 
structure (red dashed lines), and  
(b) Wannier DOS for the Ni(${x^2-y^2}$) (red) and apical Ni($s$) 
orbital (blue).  }
	\label{fig:dft}
\end{figure*}

The \textit{ab initio} electronic structure calculations using density 
functional theory (DFT) were performed with the Quantum Espresso code 
\cite{QE1,QE2,QE3} using a plane-wave pseudopotential method, 
combining a projector augmented wave method \cite{Lej16} and a specific 
choice of pseudopotentials~\cite{pot}. Within this work we chose an 
energy cutoff of $600$ eV and a $\Gamma$-centered  Brillouin zone mesh 
of $16\times16\times16$. For all calculations we used the same crystal 
structure published in Ref.~\cite{Gu20}. The two-band model is derived 
by performing a Wannier projection onto the DFT band structure as 
implemented within the Wannier90 interface~\cite{Wan90}, including 
onsite energies and hopping parameters of each Wannier orbital. The 
projection was performed onto a Ni(${x^2-y^2}$) and an apical Ni($s$)  
Wannier orbital within an energy window ranging from $-2.0$ to $4.0$ eV.  
The DFT band structure, represented here by the projected Wannier bands, 
and its density of states (DOS), are shown in Fig.~\ref{fig:dft}. Panel 
Fig. \ref{fig:dft}a shows the DFT band structure (black lines) and 
Wannier bands (red dashed lines). The right panel Fig. \ref{fig:dft}b 
contains the resulting Wannier DOS around the Fermi energy for the bands 
of two symmetries: ${x^2-y^2}$ and $s$. 
Our result qualitatively matches previous studies~\cite{Gu20,Adh20}.  

\begin{table}[b!]
	\caption{Hoppings parameters for the two-band model (all in eV)
		used in ED calculations.}
	\centering
\begin{ruledtabular}
	\begin{tabular}{cc}
		$(x,y,z)$ & $t^{\alpha\beta}_{ijk}$ 
		\\ 
		\hline
		(0,0,0) &  $\begin{pmatrix}
			0.2 & 0 \\
			0 & 1.2 
		\end{pmatrix}$ \\ 
		(1,0,0) &
		$\begin{pmatrix}
			-0.380 & -0.050 \\
			-0.050 & -0.031 
		\end{pmatrix}$  \\
		(0,1,0) &
		$\begin{pmatrix}
			-0.380 & 0.050 \\
			 0.050 & -0.031 
		\end{pmatrix}$  \\
		(0,0,1) &
		$\begin{pmatrix}
		    -0.039  & 0 \\
		     0      & -0.076
		\end{pmatrix}$  \\
		(1,1,0) & 
		$\begin{pmatrix}
			0.088 & 0 \\
			0 & -0.111 
		\end{pmatrix}$ \\
		(1,0,1) &   
		$\begin{pmatrix}
			 0.001 & -0.009 \\
			-0.009 & -0.252 
		\end{pmatrix}$\\
		(0,1,1) &   
		$\begin{pmatrix}
			 0.001 &  0.009 \\
			 0.009 & -0.252 
		\end{pmatrix}$\\
		(1,1,1) &   
		$\begin{pmatrix}
			0.015 & 0 \\
			0 & 0.056 
		\end{pmatrix}$\\
	\end{tabular}
		\label{tab3}
\end{ruledtabular}
\end{table}

The Wannier Hamiltonian is given within the real space notation and is 
of the form,
\begin{align}
\mathcal H(\mathbf R) &= \sum_{ij\alpha\sigma} 
t_{ij\alpha}(\mathbf R)c_{i\alpha\sigma}^\dagger c_{j\alpha\sigma}^{}.   
\end{align}
The basis is ordered following the convention $\{d_{x^2-y^2},s\}$. 
Explicit numerical values for the hopping parameters are given in Table 
\ref{tab3} for the terms with leading contributions, i.e., terms larger 
than $0.001$ (the other terms were neglected). Note that the relation 
$\mathcal H(-\mathbf R)=\mathcal H^T(\mathbf R)$ holds for Wannier 
models. As a consequence, both terms (for distances $\mathbf R$ and 
$-\mathbf R$) need to be included in further calculations. From  
$\mathcal H(\mathbf R)$ a tight-binding Hamiltonian can be constructed 
by applying a Fourier transformation of the form,
\begin{align}
	\mathcal H_{ij}(\mathbf k)=\sum_{\mathbf R}
	\mathrm e^{\mathrm i\mathbf k \mathbf R}\mathcal H_{ij}(\mathbf R),
\end{align}
where $\mathbf R$ describes the distances of the Wannier orbitals 
$|i-j|$ and is typically represented in terms of the lattice vectors. 

\section{Effective two-band Hamiltonian}

Following the idea of neglecting oxygen orbitals, we construct an 
effective model containing only Ni($3d$) orbitals. The band 
structure calculation shows that only two bands contribute at Fermi 
level. Therefore, the two-band model of nickelates is capable of 
reproducing the physics of nickelates. It consists of $x^2-y^2$ orbital 
and the $s$ orbital which includes rare-earth $5d$ states 
and Ni apical orbitals. 

We consider the two-band Hamiltonian $H = H_{\rm kin} + H_{\rm int}$.
The kinetic part is 
\begin{equation}
\label{twoba}
H_{\rm kin} = \sum_{i\alpha\sigma}\epsilon_{\alpha}
	a^{\dagger}_{i\alpha\sigma}a_{i\alpha\sigma}^{} 
+ \sum_{ij\alpha\beta\sigma}t_{ij}^{\alpha\beta}
a^{\dagger}_{i\alpha\sigma}a_{j\beta\sigma}^{}\,,
\end{equation}
while the interactions are given in a similar way to Eq. \eqref{int}.
%
The orbitals ${x^2-y^2}$ and $s$ have the diagonal energies 
$\epsilon_{\alpha}$, being 0 and $\epsilon$. The bands are constructed 
following Ref. \cite{Adh20}. The oxygen $2p$ orbitals are included 
implicitly in ${x^2-y^2}$ ones. Note that one should not confuse here 
$x^2-y^2$ symmetry with Ni($x^2-y^2$) orbital. The former is an 
effective orbital including in-plane oxygen while the latter is solely 
a Ni($3d$) orbital. The $s$ orbital contains of contributions from 
Nd($5d$), Ni($3d_{z^2}$) and Ni($4s$) orbitals. The symmetry of this 
hybrid orbital is the same as that of atomic $s$ orbital. 

Instead of Nd($4f$) electronic states, the empty Nd($5d$) orbitals are 
responsible for the striking electron pocket at the $\Gamma$ point in 
the band structure \cite{Bot20}. The Nd atoms are originally located 
in the off-plane direction of Ni-O plane. In this effective model the 
rare-earth atoms are included into Ni atoms via $s$ orbital, 
indicating the model is three-dimensional and the Coulomb interaction 
as well as Hund's coupling need to be screened. For instance, the two 
electrons sitting within $s$ orbital could be located either at Nd or 
at Ni atom. Therefore, the Coulomb repulsion between these two 
electrons within the $s$ orbital is reduced. We introduce a parameter 
$\alpha\in [0,1]$ to represent the reduced Coulomb interaction and 
Hund's coupling, i.e., $U_2=\alpha U_1$ and $J=\alpha J_H$. 
$\alpha =1 (0) $ stands for weak (strong) screening effect from 
rare-earth atoms, and we consider two parameter sets $A$ and $B$, given 
in Table \ref{tab2}. We study the effective two-band model via Lanczos 
algorithm \cite{Koch} on a $2\times2\times2$ unit cell. 

\begin{table}[b!]
\caption{Parameters of the two-band model (all in eV) used in exact 
diagonalization calculations. \\ The reference energies for the two 
bands of $x^2-y^2$ and $s$ symmetry are 0 and $\epsilon$, respectively.}
 \centering
\begin{ruledtabular}
	\begin{tabular}{ccccc}
	set &$\epsilon$&  $t$   & $U_1$ & $J_H$  \\
			\hline
	$A$ &   1.21   &  0.38  &  8.0  &  1.2   \\
	$B$ &   1.21   &  0.38  &  4.0  &  0.6   \\
		\end{tabular}
		\label{tab2}
\end{ruledtabular}
\end{table}

\begin{figure*}[t!]
	\centering
	\includegraphics[width=14cm]{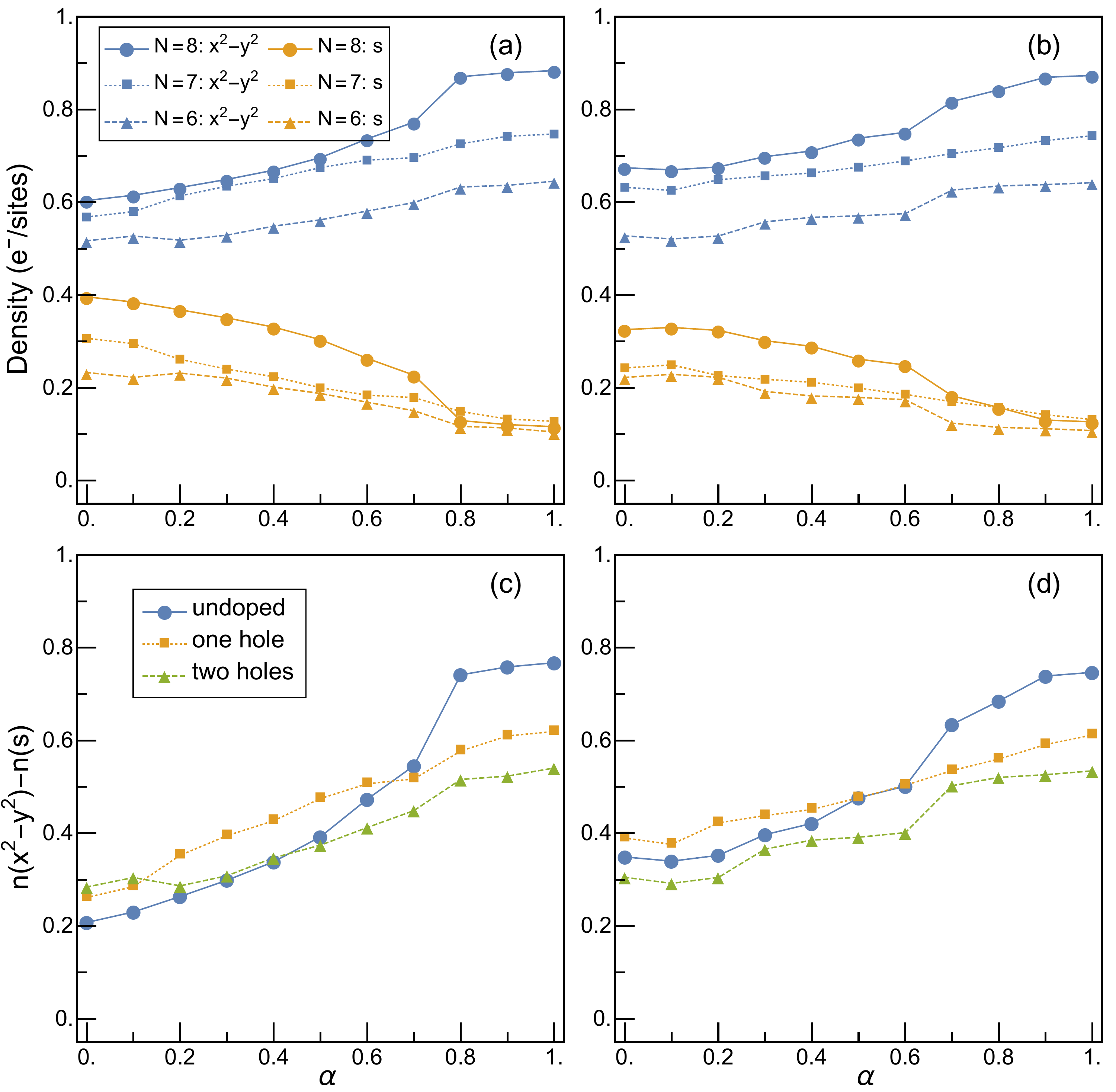}
\caption{Orbital-resolved electron densities as obtained for:	
(a) set $A$, and 
(b) set $B$, see Table \ref{tab2}. 
Strongly anisotropic electron distribution over the $d_{x^2-y^2}$ and 
$s$ orbitals is favored when the interactions are screened by 
$\alpha>0.5$, as shown for:
(c) set $A$, and 
(d) set $B$.} 
	\label{fig2}
\end{figure*}

The undoped nickelate correponds to quarter-filling, i.e., 8 electrons. 
The stoichimetric compound has $d^9$ configuration, where Ni($x^2-y^2$) 
orbital is half-filled, and the weak hybridization of Ni and Nd causes 
Ni($x^2-y^2$) orbital to be away from half-filling and creates the
self-doping effect \cite{ZYZ20,Lec21}. The Ni-Ni hopping integrals are 
obtained via fitting Wannier functions on DFT band calculation. We aim 
to address the question: where the doped holes are located in the 
two-band model? 

In the absence of electron hoppings Hamiltonian, the ground state is 
a trivial antiferromagnet where $x^2-y^2$ orbital is half-filled and 
the two canonical AFM phases, $C$-AFM and $G$-AFM, are degenerate. 
Including hopping elements to further neighbors leads to metallic 
behavior. In one-band Hubbard model the metal-insulator transition 
occurs when $U=2zt$, where $z=4$ is the number of neighbors in the 2D 
plane. Similarly, the two-band description can lead to partial 
orbital-selective Mott transition where one band is insulating and the 
other one remains metallic. In the one-band version, the two relavant 
configurations are singly occupyied or form double occupation within 
a single site. The quarter filling two-band version, however, contains 
serveral possible configurations where the lowest energy is still a 
single occupancy followed by a local triplet state with energy 
$(U-3J_H)$ \cite{Ole83}.

\section{Results and Discussion}

We begin with electronic density distribution on $2\times 2\times 2$ 
clusters obtained by exact diagonalization with twisted boundary 
condition (TBC). The hopping parameters used here are given in Table 
\ref{tab3}. With PBC, one requires hopping integrals 
$t^{\alpha\beta}_{ij}$ to be scaled by the factor of 1/2 due to 
double-counting at the boundary. To avoid this additional factor in the 
hopping terms, we replace the PBC with TBC. Instead of having a constant 
phase $t_{N+i}=t_i$ as in PBCs, the hopping terms at the boundary are 
modified by twisted angles $(\phi_{x},\phi_{y},\phi_{z})$, giving 
$t_{N+i}=e^{i\vec{\phi}\cdot\vec{r}_{i}}t_{i}$; for more details see 
Refs. \cite{Poi91,Shi97}. The PBC corresponds to $(0,0,0)$, while 
$(\pi,\pi,\pi)$ is obtained for the anti-periodic boundary condition. 
In what follows, the 
observables are obtained by averaging over several twisted angles. 

\begin{figure*}[t!]
	\centering 
	\includegraphics[width=14cm]{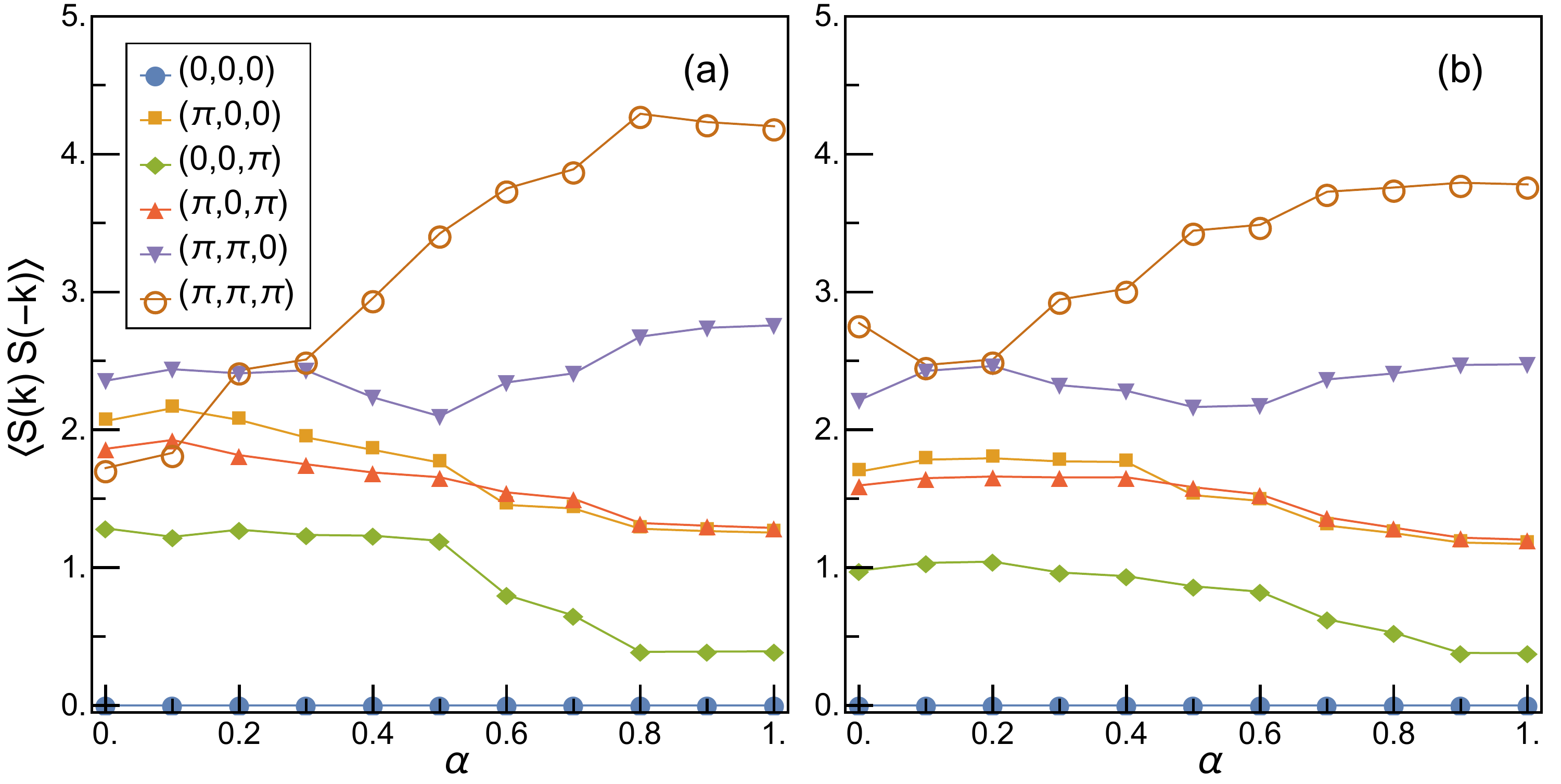}
\caption{Spin structure factor \eqref{spin} of the three-dimensional 
two-band model, as obtained for:	
(a) set $A$, and 
(b) set $B$, see Table \ref{tab2}.}
	\label{fig3}
\end{figure*}

In Figs. \ref{fig2}a-\ref{fig2}b, we show the undoped orbital-resolved 
densities, i.e., the electron occupancy within each orbital: $x^{2}-y^2$ 
(predominantly occupied) and $s$ (it usually has decent amount of 
electrons). Finite occupancy of $s$ orbital is expected due to 
self-doping effect. We further see that increasing of the parameter 
$\alpha$ causes the occupation on $x^2-y^2$ orbital to approach 
half-filling. Once the $x^2-y^2$ orbitals are almost half-filled, we 
show below that the $G$-AFM ground state is the only dominant magnetic 
ground state. Including screening effect into $x^2-y^2$ orbital by 
lowering its Coulomb interaction gives this orbital as slightly more 
favorable, see also Fig. \ref{fig2}a-\ref{fig2}b. 

By adding one hole, we see that both orbitals can be occupied by holes 
depending on how strong the screening effects of $s$ orbital are, see 
Figs. \ref{fig2}c-\ref{fig2}d. The screening effect on $x^2-y^2$ 
orbitals makes little changes in the one-hole case. An interesting 
scenerio arises when two holes are added into the undoped Ni-O plane. 
When taking the screening effect into account, holes are effectively 
occupying the $x^2-y^2$ orbital regardless of the parameter $\alpha$. 

Theoretical studies based on Ni($e_g$) bands \cite{Lec21,Tha21,ZYZ20} 
have suggested that high-spin $S=1$ state Ni$^{2+}$ is favorable for 
hole doping. In contrast, RIXS measurements \cite{Ros20} show that two 
holes are residing mainly within Ni($x^2-y^2$). Due to the limitation of 
the measurement, we cannot determine the occupation of the rare-earth 
$5d$ states. The difference between the predictions of the theoretical 
model and the experiment arises from the rare-earth atoms. In the 
Ni($e_g$) with $d^8$ configuration, the $x^2-y^2$ is half-filled and 
forming a high-spin state together with another $3d$ orbital with energy 
$(U-3J_H)$ \cite{Tha21}. This energy is significantly smaller than the 
energy $U$ for a low-spin $d^8$ state. On the contrary, the $s$ orbital, 
in the hole picture, has lower energy than $x^2-y^2$ orbital and the 
quarter-filling of electrons corresponds to $\frac{3}{4}$-filling by 
holes via particle-hole transformation. In this case, the $s$ orbital 
is filled by holes and leaves half-filled $x^2-y^2$ orbital---then holes 
reside mainly on $x^2-y^2$ orbital. 
This intuitive picture of hole configuration requires the $x^2-y^2$ to 
be nearly half-filled, in other words, it suggests that the ground 
state is a strong antiferromagnet as in cuprates \cite{Zha88,Ani99}. 
However, NMR experiments \cite{Hay03,Hay99} report no observation of 
long-range AFM order in the $R$NiO$_2$ ($R$=Nd,La) down to 2 K.

\begin{figure*}[t!]
	\centering
	\includegraphics[width=14cm]{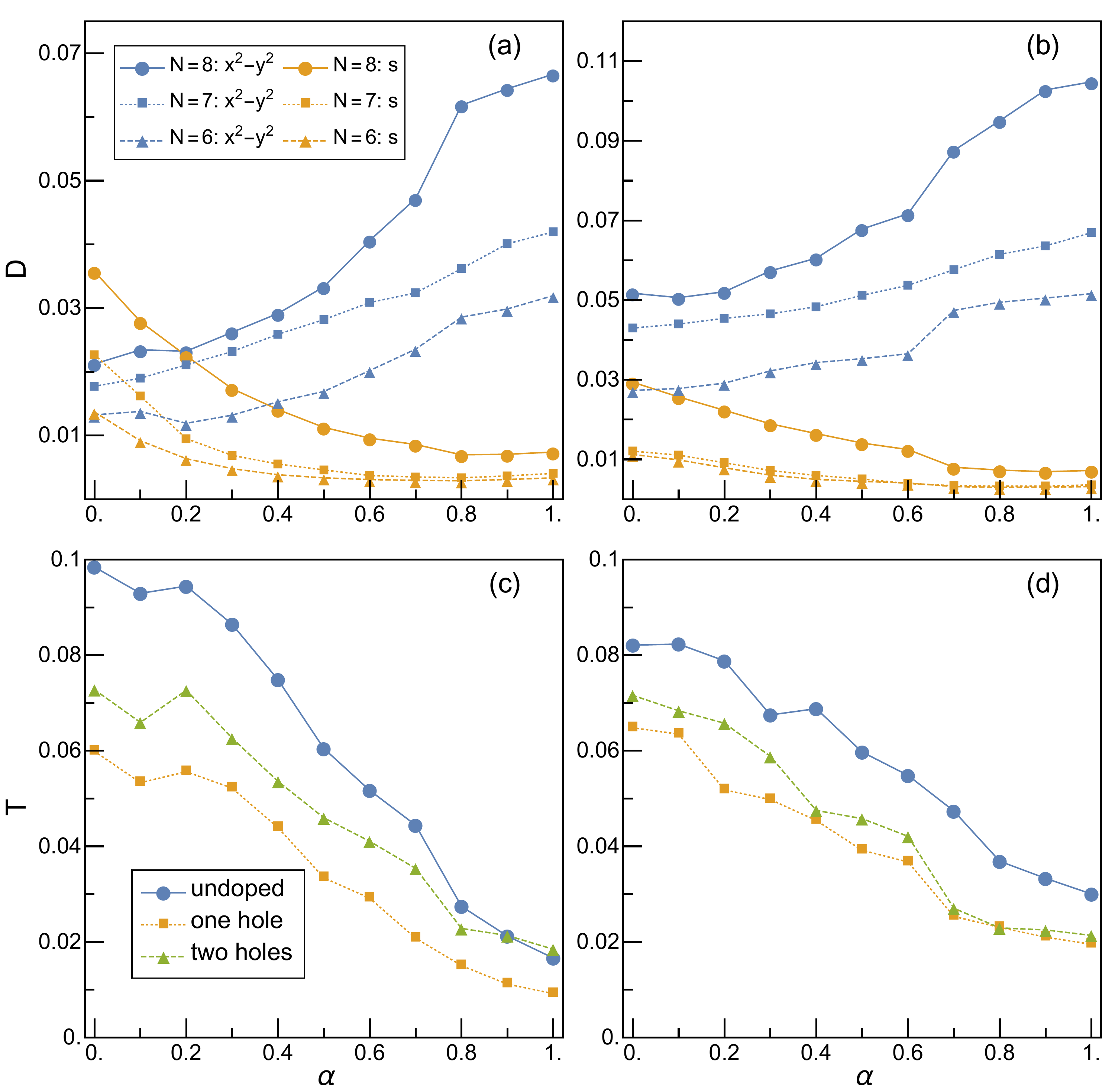}
\caption{Competition between low-spin and high-spin states as obtained 
for the parameters of:
\mbox{(a,c)~set~$A$,} and 
(b,d)~set $B$, see Table \ref{tab2}. 
Panels (a) and (b) show local double occupancy $D_{\alpha}$ in each 
$\alpha=x^2-y^2,s$ orbital, while panels (c) and (d) show local triplet 
states $T$.
}
	\label{fig4}
\end{figure*}

According to DFT+sicDMFT approach \cite{Lec21}, the paramagnetic ground 
state has the lowest energy followed by $C$-AFM with energy difference 
about 20 meV/atom and $G$-AFM with energy 105 meV/atom. The $C$-AF state 
has parallel spins along the $c$ axis, while $G$-AFM phase has 
antiparallel spin alignment to its nearest neighbor in all directions. 
To address the magnetic order in this two-band model, the spin structure 
factor of undoped profile, 
\begin{equation}
\label{spin}
\langle S(k)S(-k)\rangle=\sum_{ij}e^{i\vec{k}(\vec{r}_i-\vec{r}_j)}
\left\langle\vec{S}_i\cdot\vec{S}_j\right\rangle,
\end{equation}
is shown in Fig. \ref{fig3}. With this small cluster size we can observe 
only six relevant ${\bf k}$-points in the Brillouin zone. The spin 
structure factor at $(\pi,\pi,\pi)$ is enhanced as the $x^2-y^2$ is 
close to half-filled band, indicating that $G$-AFM 
order is favored, see Fig. \ref{fig3}. 
At small $\alpha$, the spin structure factor \eqref{spin} shows a 
competition between several magnetic ordered states, dominated by 
$(\pi,\pi,0)$ and $(\pi,\pi,\pi)$. In the recent $x$-ray scattering 
experiment \cite{Lu213}, the existence of AFM correlations was confirmed 
but the small cluster size prevents us from concluding whether AFM 
long-range order could be stable in this regime.

The screening effect from rare-earth atoms \cite{Sak20,ZYZ20} could be 
responsible for the competition of magnetic orders. At small $\alpha$ 
where the screening effect from rare-earth is strong, the electrons tend 
to form on-site triplets with energy $\epsilon+\alpha(U_1-3J_H)$ 
\cite{Tha21}, coexisting with singly occupied $x^2-y^2$ orbitals. Then 
increasing $\alpha$ enhances the interactions on $s$ orbitals and the 
on-site triplet energy surpasses the bandwidth of $s$ orbital, i.e., 
$\epsilon+\alpha(U_1-3J_H)>2zt$ ($z=4$). The largest hopping elements 
are the in-plane hoppings along $x$ and $y$ direction of $x^2-y^2$ 
orbital. Another transition then occurs when the on-site triplet energy 
overcome the bandwidth of $x^2-y^2$ orbital, becoming a Mott insulator. 

Figure \ref{fig4} shows the competition between low-spin and high-spin 
states in NiO$_2$ planes of the novel Ni-layered superconductors for 
decreasing number of holes, i.e., for electron doping. Here we use
$D_\alpha=\frac{1}{N}\sum_i 
\left\langle n_{i\alpha\uparrow}n_{i\alpha\downarrow}\right\rangle$ and
$T=\frac{1}{N}\sum_i \left\langle\frac34+
\vec{S}_{i1}\cdot\vec{S}_{i2}\right\rangle$,
where $N$ is the number of lattice sites. We~show here that the double 
occupancy $D_\alpha$ is reduced with decreasing number of holes in the 
plane. Similarly, the amplitude of high-spin states $T$ at Ni sites is 
reduced in this regime. So we conclude that electron doped 
materials have low-spin configuration. On the contrary, hole doping may 
favor locally high-spin $(S=1)$ states instead of singlet $(S=0)$ 
double occupancies of $x^2-y^2$ orbitals \cite{Tha21}.

\section{Summary and conclusions}

In summary, we have replaced the half-filled $d-p$ charge-transfer model 
by the effective two-band model at quarter-filling. The characters of 
each band are given by ${x^2-y^2}$ and $s$ symmetry. The contributions 
of Nd and Ni atoms to $s$ orbital lead to the reduction of Hund's 
exchange and Coulomb repulsion. Its strength is scaled by the parameter 
$\alpha$. The model shows the three competing phases: metal, 
orbital-selective and Mott insulator. The Mott insulator is realized 
when the splitting between on-site triplet and singly occupied state is 
larger than the size of the bandwidth, similar to one-band Hubbard 
model. It is followed by the orbital-selective Mott insulator where the 
orbital splitting separates the $s$ and $x^2-y^2$ bandwidth. 

The nonmagnetic ordering is unlikely within this small unit cell and 
only AFM configuration can be realized. While the $G$-AFM phase is 
clearly dominating at large $\alpha$, the competition between $C$-AFM 
and $G$-AFM is found at low $\alpha$ where the on-site triplet competes 
with AFM ground state, indicating the tendency toward AFM ordering. The 
holes are doped differently among the three phases. In metallic phase, 
the itinerant $s$ orbital is favorable for holes. On the other hand, in 
the orbital-selective phase, the situation is slightly complicated while 
adding one hole favors $s$ orbital but adding two holes can effectively 
lead again to $x^2-y^2$ occupation, as in the Mott insulating phase. 

When the screening effect on $x^2-y^2$ orbital is included, the 
stability of orbital-selective phase is enhanced and metallic phase 
vanishes. The two-band model at quarter-filling, therefore, connects the 
two controversial scenerios of which orbitals preferred by doped-holes 
via the parameter $\alpha$. The parameter somehow represents the 
screening effect from rare-earth orbital, showing the importance of 
rare-earth atoms for the electronic structure of superconducting 
nickelate.

\textbf{Authors' Contributions:} 
Tharathep Plienbumrung perfomed numerical analysis of the two-band 
model. The two-band model was derived from the $d-p$ charge-transfer  
model by Michael Schmid. All authors selected the relevant information, 
analyzed the predictions of the theory versus experimental data, 
developed the interpretation of the numerical results, and wrote the 
manuscript.

\textbf{Funding:} T.~P. acknowledges \mbox{Development} 
and Promotion of Science and Technology Talents Project (DPST). 
A.~M.~Ole\'s kindly acknowledges Narodowe Centrum Nauki
(NCN, National Science Centre, Poland) Project No. 2016/23/B/ST3/00839.

\textbf{Acknowledgments:} 
We would like to thank Andres Greco, Peter Horsch, Krzysztof 
Ro\'sciszewski, and George A. Sawatzky for many insightful discussions. 
A.~M.~Ole\'s is grateful for the Alexander von Humboldt Foundation
Fellowship (Humboldt-Forschungspreis).

\textbf{Conflicts of Interest:} We declare no conflicts of interest.


\end{document}